\def\breakon{\end{multicols}\widetext\vspace{-.6cm}
\noindent\rule{.49\linewidth}{.3mm}\rule{.3mm}{.5cm}\vspace{0.0cm}}
\def\breakoff{\vspace{-0.45cm}
\noindent
\rule{.50\linewidth}{.0mm}\rule[-.47cm]{.3mm}{.5cm}\rule{.49\linewidth}{.3mm}
\vspace{-0.25cm}
\begin{multicols}{2}   }
\documentstyle[aps,epsf,prb,multicol]{revtex}
\begin{document}

\makeatletter
\renewenvironment{table}
  {\let\@capwidth\linewidth\def\@captype{table}}
  {}

\renewenvironment{figure}
  {\let\@capwidth\linewidth\def\@captype{figure}}
  {}
\makeatother

\title{Phase Transitions Between Topologically Distinct Gapped Phases
in Isotropic Spin Ladders}
\author{Eugene H. Kim,$^\dag$ G. F\'ath,$^\ddag$ 
        J. S\'{o}lyom,$^\ddag$ and D.~J. Scalapino$^\dag$}
\address{\dag Department of Physics, University of California,  
          Santa Barbara, California 93106-9530 \\
         \ddag Research Institute for Solid State Physics, P.O. Box 49,
          H-1525 Budapest, Hungary}          
\maketitle

\begin{abstract}
We consider various two-leg ladder models exhibiting gapped phases.
All of these phases have short-ranged valence bond ground states,
and they all exhibit string order.  However, we show that short-ranged
valence bond ground states divide into two topologically distinct
classes, and as a consequence, there exist two topologically distinct 
types of string order.  Therefore, not all gapped phases belong to the
same universality class.  We show that phase transitions occur when we
interpolate between models belonging to different topological classes,
and we study the nature of these transitions.
\end{abstract}


\vspace{.15in}
\begin{multicols}{2}

\section{Introduction}

Even though the spectrum of the spin-1/2 Heisenberg chain was obtained by
Bethe\cite{bethe} almost seventy years ago using his famous Ansatz,
low-dimensional spin systems are still a strong area of activity, full
of surprises and puzzles. A major source of this activity was
Haldane's conjecture,\cite{haldane} which predicted that isotropic
antiferromagnetic Heisenberg chains with integer spin have a gapped
spectrum, while chains with half-integer spin have a gapless spectrum.  
There has been considerable theoretical and experimental evidence in
support of Haldane's conjecture; it is probably appropriate to call it a
theorem, despite the lack of a rigorous mathematical proof.\cite{sierra}
Incidentally, the gapped phase in integer spin Heisenberg chains has come
to be known as the Haldane phase.

A breakthrough in understanding the nature of the Haldane phase
came when it was realized that one can go without a phase transition from 
the spin-1 Heisenberg chain to the AKLT model,\cite{aklt} where the 
ground state is made up solely of nearest neighbor valence bonds.
The Haldane gap is thus related to the energy needed to break
short-ranged valence bonds.

\vspace{.2in}
\begin{figure}
\epsfxsize=3.25in
\centerline{\epsfbox{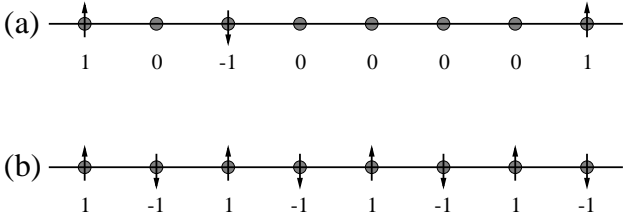} }
\vspace{.2in}
\caption{(a) Typical configuration of the spin-1 chain.  (b) N\'{e}el
 order after removing all sites with $S^{\alpha}_i = 0$.}
\label{fig:string}
\end{figure}
\vspace{.2in}

Another important step was when den Nijs and Rommelse identified a 
hidden order in the Haldane phase of the spin-1 chain.\cite{dennijs}
They showed that although the sites with 
$S^{\alpha}_i = 1, 0, -1$ are not well ordered in position, their 
sequence is ordered in the way shown schematically in 
Fig.~\ref{fig:string}.  That is, if we remove all sites with 
$S^{\alpha}_i = 0$, the remaining sites have N\'eel order.  The order
parameter which reveals this hidden order is the non-local string
order parameter
\begin{equation}
  {\cal O}^{\alpha} = - \lim_{|i-j| \rightarrow \infty}  
        \left\langle S^{\alpha}_i 
        \exp\left( i\pi \sum_{l=i+1}^{j-1} S^{\alpha}_l\right) 
        S^{\alpha}_j \right\rangle \, ,
\end{equation}
where $S^{\alpha}_i$ is the spin-1 operator at site $i$, and 
$\alpha = x,y,z$.

A further impetus for the study of low-dimensional spin systems
was given recently by the discovery of spin-ladder materials.\cite{rice}  
Since the spin-1/2 Heisenberg chain has a gapless excitation spectrum 
with spin-spin correlation functions exhibiting power-law behavior, it 
initially came as a surprise when it was found that 
the two-leg ladder had a gapped spectrum with exponentially decaying 
spin-spin correlations,\cite{dagotto} while the gapless spectrum survived 
in the three-leg ladder.  Thus ladders could have a gapped or gapless 
spectrum, depending on the number of legs.  More specifically, Heisenberg 
ladders with an even number of legs have a gapped spectrum, while ladders 
with an odd number of legs have a gapless spectrum. 

The appearance of a gapped spectrum for even-legged ladders 
and a gapless spectrum for odd-legged ladders is highly reminiscent of 
Haldane's conjecture for spin chains.  Therefore, a natural question 
arises: Is the gapped phase in spin ladders related to the Haldane phase 
in spin chains? In particular, does the gapped phase exhibit string order?  
For the rest of the paper, we will restrict our discussion to the two-leg 
ladder.

For the case of ferromagnetic interchain coupling, it is clear that the
two-leg ladder can be equivalent to the spin-1 chain, with the two
spins on a rung forming an effective $S=1$.  Such a phase is directly
related to the Haldane phase, and the system has string order. Similarly,
if the interchain coupling is antiferromagnetic and along plaquette
diagonals, when the interchain coupling is equal to the coupling along
the chains, the model is in fact the composite spin representation of the 
spin-1 chain; thus, the gapped phase is equivalent to the Haldane phase. 
(See Sec. II for a more detailed discussion of the composite spin 
representation.)  Besides the gap, these models exhibit another 
characteristic feature of the Haldane phase. Namely, the ground state is 
unique if periodic boundary conditions are used, but it is four-fold 
degenerate for open boundary conditions.

A gap appears in the excitation spectrum of spin ladders for other types
of interchain coupling as well, namely for antiferromagnetic coupling
along the rungs or ferromagnetic coupling along plaquette diagonals.
Whether these gapped phases are related to the Haldane phase is
much less obvious. For example, for antiferromagnetic interchain coupling
along the rungs, the zeroth order picture is given by studying the limit
in which the interchain coupling is much larger than the coupling along
the chains (i.e., $J_{\perp} \gg J$).  In this limit, the ground state is
well described by a product of rung singlets, with an energy gap $\sim
J_{\perp}$ to break a singlet and form a triplet excitation.  Here, the 
ground state is unique, irrespective of whether open or periodic boundary
conditions are used. Nevertheless, as was demostrated by
White,\cite{white} the antiferromagnetic ladder can be transformed 
continuously to a model (seemingly) equivalent to the composite spin 
representation of the spin-1 chain by switching on an irrelevant further 
neighbor coupling.  Consequently, the antiferromagnetic ladder is also
related, in some way, to the Haldane phase and has string
order.\cite{white,nishiyama}

There are also other types of ladders exhibiting spin gapped phases, so
the question of string order in ladders and the relationship of the gapped
phases to the Haldane phase becomes even more interesting.  In particular,
the spin-1/2 chain with second neighbor coupling is often represented as a
two-leg zig-zag ladder.\cite{wata-yoko} For a particular value of the
couplings, the so-called Majumdar-Ghosh point,\cite{majumdar} the ground
state is known exactly.  It is doubly degenerate and the excitation
spectrum is gapped; each ground state consists of a sequence of
independent singlets.  However, it was shown that the Majumdar-Ghosh
ground state has perfect string order.\cite{hida}
It has also been shown\cite{mikeska} that the Majumdar-Ghosh model
can be smoothly connected to the ladder with strong ferromagnetic rung
coupling, which is equivalent to a spin-1 chain, without a phase 
transition.  

Since the ground state of the Majumdar-Ghosh model consists 
of decoupled singlets, similar to the ground state of the ladder with 
strong antiferromagnetic coupling along the rungs, one might get the 
impression that all spin-gapped phases can be smoothly connected to 
each other.  By this we mean that one can go from one model to another
by continuously varying the model parameters without undergoing a phase 
transition. In this paper, we show that the gapped phases in 
isotropic two-leg spin ladders divide into two topologically distinct
classes.  This implies that phase transitions must necessarily occur if 
we try to interpolate between models belonging to different topological 
classes.  Although not all gapped phases are directly equivalent to the 
Haldane phase of the spin-1 chain, all possess some kind of string order.

The rest of the paper is organized as follows.  In Sec.~II we introduce
the spin ladder models that we will consider, and we recapitulate briefly
what is known about these models.  In Sec.~III we discuss the relationship 
between valence bond states and string order, and also the possibility 
of phase transitions between spin models with topologically different 
string order.  A bosonization treatment of the various models is presented 
in Sec.~IV, and a discussion of the results obtained is given in Sec.~V. 
Finally, Sec.~VI gives a summary of our results and an outlook for
future work.


\section{The Models}

We begin with two antiferromagnetic spin-1/2 Heisenberg chains with 
the Hamiltonian
\begin{equation}
    {\cal H}_0 = J \sum_{i=1}^N \left( 
     \vec S_{i,1} \cdot \vec S_{i+1,1} + \vec S_{i,2} \cdot \vec S_{i+1,2}
    \right),
\end{equation} 
where $\vec S_{i,1}$ ($\vec S_{i,2}$) is the spin operator at site $i$
on chain 1 (chain 2).  We will consider various forms for the interchain 
coupling.  

\begin{figure}
\epsfxsize=3.25in
\centerline{\epsfbox{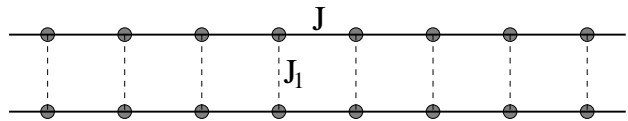} }
\vspace{.2in}
\caption{The usual two-leg ladder.}
\label{fig:ladder}
\end{figure}

The interchain coupling 
\begin{equation}
    {\cal H}_1 = J_1 \sum_{i=1}^N \vec S_{i,1} \cdot \vec S_{i,2}
\end{equation} 
describes the usual rung coupling along the legs of the ladder.  This type
of ladder is shown in Fig.~\ref{fig:ladder}.  (We will simply refer 
to this ladder model as a {\sl ladder}.)  

\vspace{.2in}
\vspace{.2in}

When $J_1$ is strongly ferromagnetic ($J_1 < 0$ and $|J_1| \gg J$) the 
two spins on the rung form a triplet, the singlet being much higher in 
energy. In this limit, the ladder behaves like a spin-1 chain, and hence the 
spectrum is gapped.  However, it has been shown\cite{barnes} that a gap is 
generated by an arbitrarily small ferromagnetic coupling. Therefore, it 
appears that weak and strong ferromagnetic coupling are continuously related.

\vspace{.2in}
\begin{figure}
\epsfxsize=3.25in
\centerline{\epsfbox{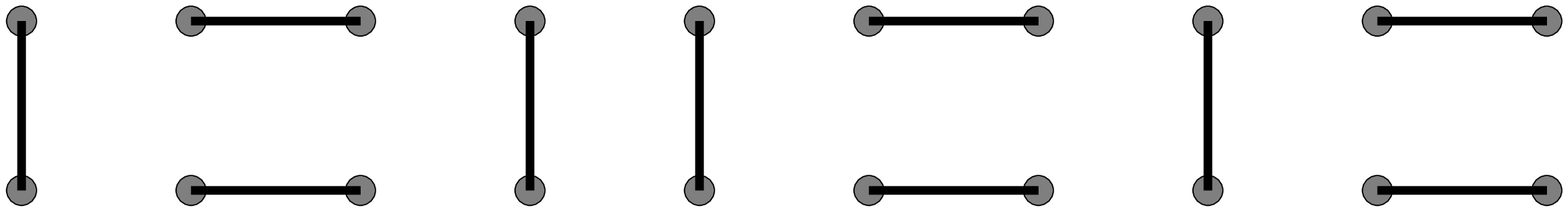} }
\vspace{.2in}
\caption{A typical configuration in the RVB state of the antiferromagnetic
ladder.  Solid lines represent singlet bonds.}
\label{fig:configurationrvb}
\end{figure}
\vspace{.2in}

When $J_1$ is strongly antiferromagnetic ($J_1 > 0$ and $J_1 \gg J$),
the ground state is essentially a product of rung singlets with a 
gap to magnon excitations.  When $J_1 = J$, it was shown that the 
ground state is well described by a nearest neighbor resonating valence 
bond (RVB) state and has a gap to the excited states.\cite{rvb}    
A typical configuration of the RVB state is shown in  
Fig.~\ref{fig:configurationrvb}. Similar to the ferromagnetic case,
it has been shown\cite{watanabe,barnes2} that the spectrum is gapped 
for arbitrarily small antiferromagnetic interchain coupling.  
Therefore, weak and strong antiferromagnetic coupling also seem to be 
continuously related.

\begin{figure}
\epsfxsize=3.25in
\centerline{\epsfbox{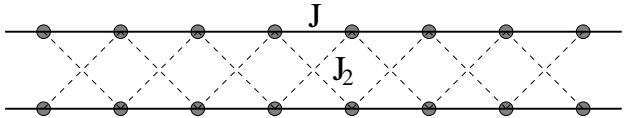} }
\vspace{.2in}
\caption{The diagonal ladder model.}
\label{fig:ladderdiagonal}
\end{figure}

We will also consider a ladder in which the interchain coupling is 
along plaquette diagonals
\begin{equation}
    {\cal H}_2 = J_2 \sum_{i=1}^N \left[ \vec S_{i,1} \cdot \vec S_{i+1,2}
      +\vec S_{i,2} \cdot \vec S_{i+1,1} \right].
\end{equation} 
Together with ${\cal H}_0$ this gives the ladder shown in 
Fig.~\ref{fig:ladderdiagonal}.  (We will refer to this ladder as a 
{\sl diagonal ladder}.)

\vspace{.2in}
\vspace{.2in}

\begin{figure}
\epsfxsize=3.25in
\centerline{\epsfbox{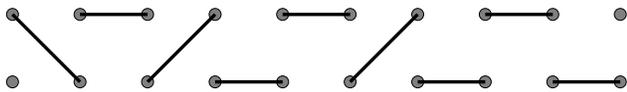} }
\vspace{.2in}
\caption{A typical configuration in the AKLT state of the diagonal
ladder.  Solid lines represent singlet bonds.}
\label{fig:configurationaklt}
\end{figure}

For $J_2 = J$, this model is in fact the {\em composite spin 
representation} of a spin-1 chain.  More precisely, by starting 
with the Hamiltonian of a spin-1 chain 
\[
    {\cal H} = J \sum_{i=1}^N \vec S_{i} \cdot \vec S_{i+1} \, , 
\]
and representing the spin-1 operator on site $i$ as the sum of 
two spin-1/2 operators, $ \vec S_i = \vec S_{i,1} + \vec S_{i,2}$,  
we find
\[
 {\cal H} = {\cal H}_0 + {\cal H}_2 \,,
\]
with $J_2 = J$.  In the composite spin representation, the total spin of 
each rung commutes with the Hamiltonian, so the eigenstates can all be 
classified by the total spin on each rung.  The set of eigenstates with 
only triplets on all of the rungs corresponds to the spectrum of the 
spin-1 Heisenberg chain. Hence, the low energy spectrum of the composite 
spin representation is identical to that of a spin-1 chain.\cite{composite} 
The ground state of this model is well described by the AKLT state,
\cite{aklt} a typical configuration of which is shown in 
Fig.~\ref{fig:configurationaklt}.  Here again, it has been 
shown\cite{solyom} that a gap appears for arbitrarily small $J_2 > 0$.
It has also been shown\cite{solyom} that the spectrum is gapped 
for $J_2 < 0$; in fact, the gap is generated for arbitrarily 
small $J_2 < 0$.  Therefore, it appears that weak and intermediate 
coupling are continuously related for both $J_2 > 0$ and $J_2 < 0$, 
however the gap vanishes at $J_2=0$.

\vspace{.2in}
\vspace{.2in}

As one can see from Fig.~\ref{fig:configurationaklt}, for a finite system
there are effectively free spin-1/2's at the ends of the ladder, which are
responsible for a four-fold degenerate ground state. When periodic
boundary conditions are used, all of the spins are bound into singlets and
the ground state is unique.

\vspace{.2in}
\begin{figure}
\epsfxsize=3.25in
\centerline{\epsfbox{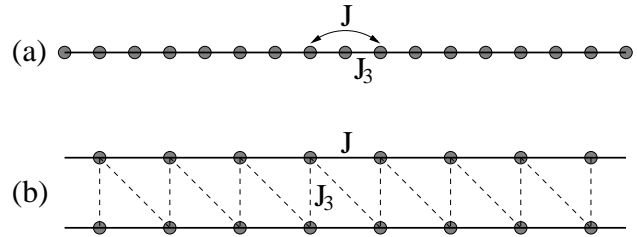} }
\vspace{.2in}
\caption{The zig-zag ladder shown as (a) a chain with first and second 
neighbor interactions (b) a ladder.}
\label{fig:ladderzigzag}
\end{figure}
\vspace{.2in}

Finally, we will consider an interchain coupling similar to ${\cal H}_2$, 
but with only one of the diagonal couplings
\begin{equation}
    {\cal H}_3 = J_3 \sum_{i=1}^N \vec S_{i+1,1} \cdot \vec S_{i,2}.
\end{equation} 
We consider such an interchain coupling because, for $J_1=J_3$, we can
write this model as a spin-1/2 Heisenberg chain with first and second
neighbor interactions
\begin{eqnarray*}
    {\cal H} & = & {\cal H}_0 + {\cal H}_1 + {\cal H}_3  \nonumber \\
             & = & J_3 \sum_{i=1}^{2N} \vec S_{i} \cdot \vec S_{i+1} +
      J \sum_{i=1}^{2N} \vec S_{i} \cdot \vec S_{i+2}   
\end{eqnarray*} 
(See Fig.~\ref{fig:ladderzigzag}).  (We will refer to this type of ladder 
[for $J_1 = J_3$] as a {\sl zig-zag ladder}.)

\vspace{.2in}
\begin{figure}
\epsfxsize=3.25in
\centerline{\epsfbox{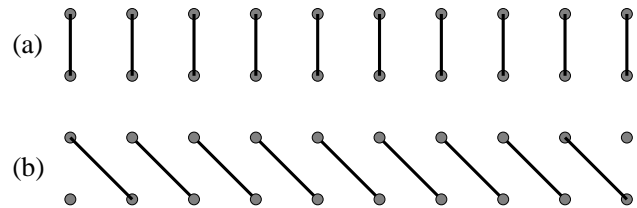} }
\vspace{.2in}
\caption{The two degenerate ground states of the zig-zag ladder at the 
Majumdar-Ghosh point.  Solid lines represent singlet bonds.}
\label{fig:majumdar}
\end{figure}
\vspace{.2in}

The zig-zag ladder is known\cite{wata-yoko} to have a quantum phase 
transition from critical, gapless behavior to a spontaneously dimerized 
gapped phase as $J$ increases at $J = 0.241 J_3$. Moreover, at the special
Majumdar-Ghosh point\cite{majumdar} ($J = 0.5 J_3$), the ground state is 
known exactly.  It is two-fold degenerate in the thermodynamic limit, and 
each of the ground states is a sequence of decoupled singlets, as shown in 
Fig.~\ref{fig:majumdar}.  In fact, it
has been shown that for $0.241 < J/J_3 < \infty$, the zig-zag ladder
is gapped and has dimer order.\cite{wata-yoko,whiteaffleck}  Therefore, it 
appears that the entire range $0.241 < J/J_3 < \infty$ is continuously 
related to the Majumdar-Ghosh point.


\section{Valence Bond States and String Order}

As previously mentioned, all of our models have short-ranged (SR) 
valence bond (VB) ground states, and all of our models exhibit string
order.  In this section, we will argue that there are two topologically
distinct types of string order and that these two types of string 
order are intimately related to the VB structure.

In general, any singlet state of an SU(2) symmetric model can be
represented in terms of VB's. The SR-VB ground states of gapped spin 
liquids are, in the typical case, a linear combination of a large number
of VB configurations, in which the probability to find a longer ranged
VB is exponentially small.  The gap in the spectrum is related to the 
finite energy needed to break a VB.  On the other hand, systems with a
gapless spectrum necessarily contain longer VB's as well.

\begin{figure}
\epsfxsize=3.25in
\centerline{\epsfbox{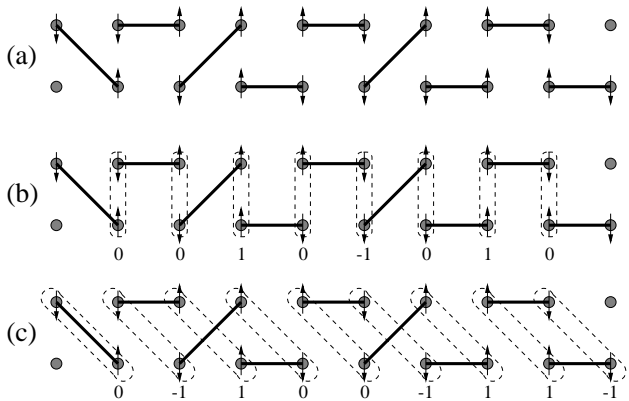} }
\vspace{.2in}
\caption{(a) One particular spin configuration of 
Fig.~\ref{fig:configurationaklt}.  
(b) Dashed lines encircling rungs: $S^z_i = S^z_{i,1} + S^z_{i,2}$.
Notice that after we remove all sites with $S^z_i = 0$, the remaining 
sites have N\'{e}el order.    
(c) Dashed lines encircling diagonals: $S^z_i = S^z_{i+1,1} + S^z_{i,2}$.  
(Figure adopted from Ref.~\ref{nishiyama}.) }
\label{fig:aklt}
\end{figure}

In order to see the connection between the VB structure and string order, 
we first consider the diagonal ladder with $J_2 = J$.  The ground state 
is well described by the RVB picture of the AKLT state.  A typical
configuration was shown in Fig.~\ref{fig:configurationaklt}. One
particular spin configuration of Fig.~\ref{fig:configurationaklt} is shown
in Fig.~\ref{fig:aklt}(a). 
Suppose we add the $z$ component of the spins on the same rung, as shown
in Fig.~\ref{fig:aklt}(b).  The total $S^z_i$ can take on the values
$1,0,-1$.  Considering this sequence, if we remove all sites with
$S^z_i=0$, the remaining sites have N\'eel order (i.e. there is string
order).  If, on the other hand, the $z$ components of the spins
along plaquette diagonals are added, as shown in Fig.~\ref{fig:aklt}(c),
there is no string order.

\vspace{.2in}
\vspace{.2in}

\begin{figure}
\epsfxsize=3.25in
\centerline{\epsfbox{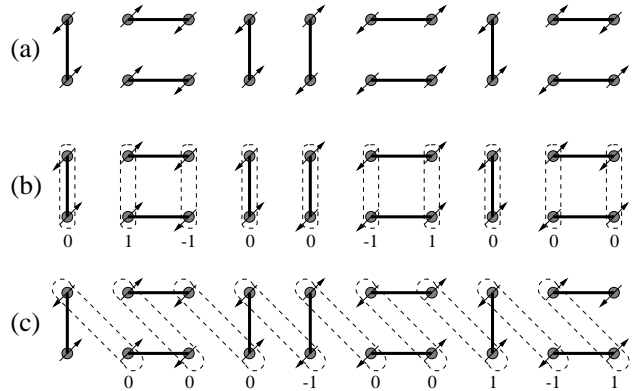} }
\vspace{.2in}
\caption{(a) One particular spin configuration of 
Fig.~\ref{fig:configurationrvb}.  
(b) Dashed lines encircling rungs: $S^z_i = S^z_{i,1} + S^z_{i,2}$.
(c) Dashed lines encircling plaquette diagonals: 
$S^z_i = S^z_{i+1,1} + S^z_{i,2}$.  
Notice that after we remove all sites with $S^z_i = 0$, the remaining
sites have N\'{e}el order.  
(Figure adopted from Ref.~\ref{nishiyama}.) }
\label{fig:rvb}
\end{figure}

Now consider the antiferromagnetic ladder with $J_1 = J$.  The ground
state is well described by a nearest-neighbor RVB state, for which a
typical configuration was shown in Fig.~\ref{fig:configurationrvb}.  One
particular spin configuration is shown in Fig.~\ref{fig:rvb}(a).  
Suppose we look at the $z$ component of the total spin on a rung, as 
shown in Fig.~\ref{fig:rvb}(b); the state has no string order.  If, 
however, we consider the $z$ component of the total spin along plaquette 
diagonals, as shown in Fig.~\ref{fig:rvb}(c), string order is found. It 
was shown by White\cite{white} that there is a $96.2\%$ probability of
finding triplets along plaquette diagonals, and it was verified 
numerically\cite{white,nishiyama} that these triplets in fact 
exhibit string order.  It was also shown\cite{white,nishiyama} that no
string order is found if the total spin along rungs is considered.

\vspace{.2in}
\vspace{.2in}

As mentioned before, the ladder with ferromagnetic coupling
along the rungs is continuously related to the true spin-1 chain.
Therefore, it has an AKLT-like ground state and string order due to
triplets along the rungs. On the other hand, the ground state of
the diagonal ladder with ferromagnetic interchain coupling has an 
RVB-like ground state, similar to that of the ladder with 
antiferromagnetic interchain coupling; as discussed above string
order is due to triplets along plaquette diagonals.

\begin{figure}
\epsfxsize=3.25in
\centerline{\epsfbox{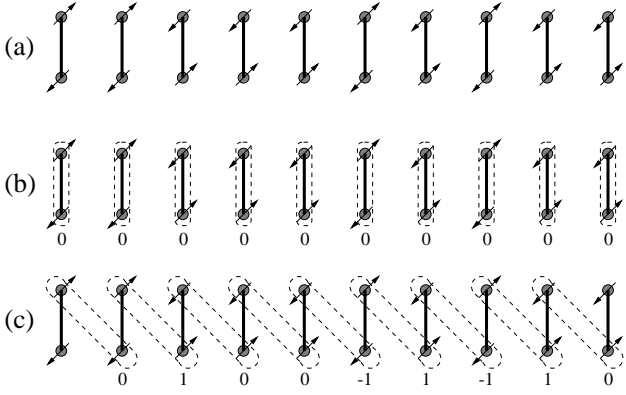} }
\vspace{.2in}
\caption{(a) One particular spin 
configuration of Fig.~\ref{fig:majumdar}(a).  
(b) Dashed lines encircling rungs: $S^z_i = S^z_{1,i} + S^z_{i,2}$.
(c) Dashed lines encircling plaquette diagonals:
$S^z_i = S^z_{i+1,1} + S^z_{i,2}$.  Notice that after we remove all sites
with $S^z_i = 0$, the remaining sites have N\'{e}el order.}
\label{fig:majumdardimer1}
\end{figure}

\begin{figure}
\epsfxsize=3.25in
\centerline{\epsfbox{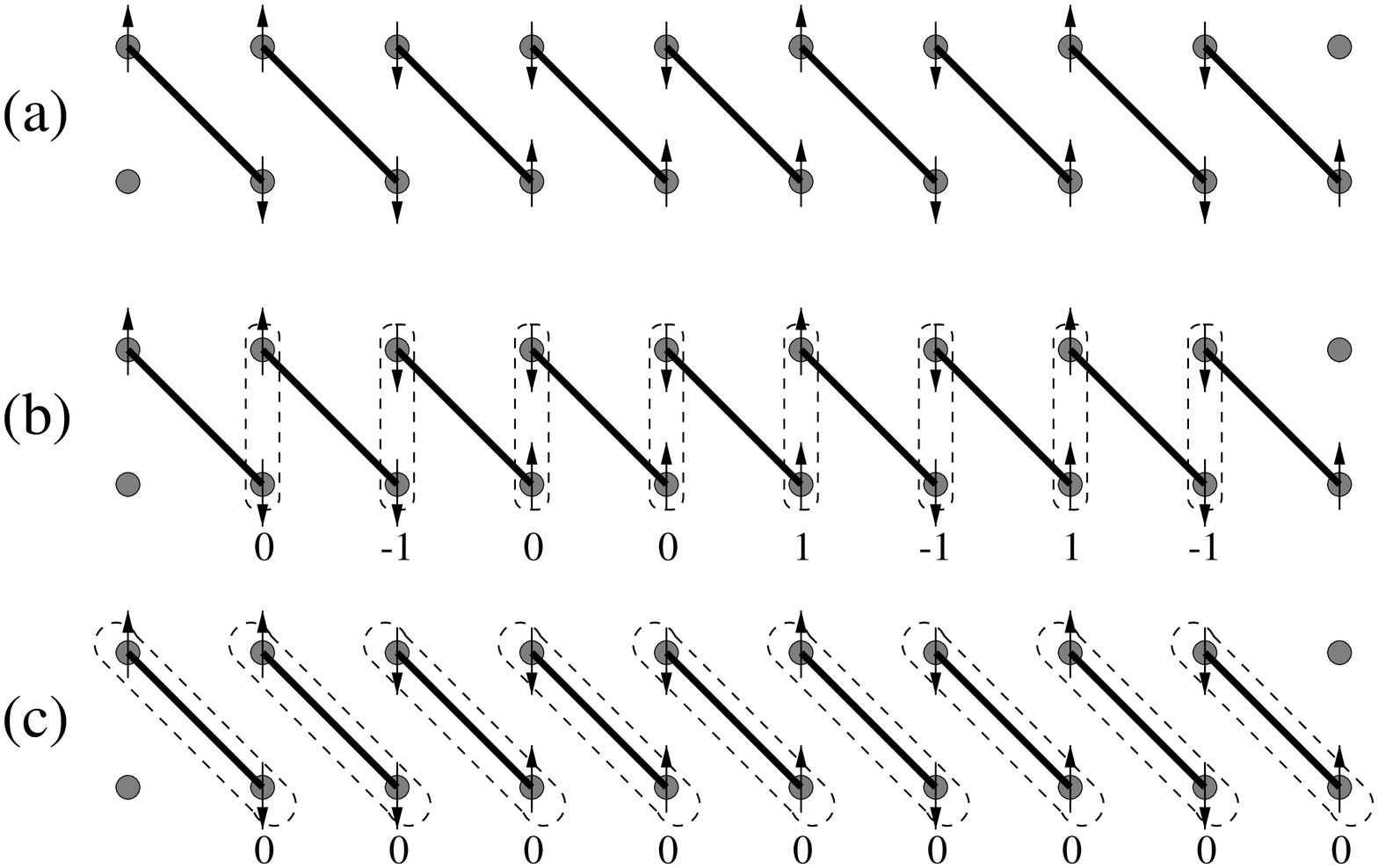} }
\vspace{.2in}
\caption{(a) One particular spin 
configuration of Fig.~\ref{fig:majumdar}(b).  
(b) Dashed lines encircling rungs: $S^z_i = S^z_{1,i} + S^z_{i,2}$.  
Notice that after we remove all sites with $S^z_i = 0$, the remaining
sites have N\'{e}el order.  (c)  Dashed lines encircling plaquette 
diagonals: $S^z_i = S^z_{i+1,1} + S^z_{i,2}$. }
\label{fig:majumdardimer2}
\end{figure}

Finally, consider the zig-zag ladder at the Majumdar-Ghosh point.  The
exact ground state is given by a product of decoupled singlets, as shown 
in Fig.~\ref{fig:majumdar}.  One particular spin configuration for 
Fig.~\ref{fig:majumdar}(a) is shown in Fig.~\ref{fig:majumdardimer1}(a).   
Adding the $z$ components of the spins on the same rung, as shown in 
Fig.~\ref{fig:majumdardimer1}(b), it is obvious that we always get zero.  
However, if we add the spins along plaquette diagonals, as shown in 
Fig.~\ref{fig:majumdardimer1}(c), we find string order.
Now, however, consider the ground state shown in Fig.~\ref{fig:majumdar}(b).  
One particular spin configuration is shown in 
Fig.~\ref{fig:majumdardimer2}(a).  Then, as shown in  
Fig.~\ref{fig:majumdardimer2}(b), if we add the spins along the
rungs, we find string order.  However, if we add the spins along plaquette 
diagonals, as shown in Fig.~\ref{fig:majumdardimer2}(c), we always get zero. 
This result is not surprising, since it was already shown\cite{hida} 
that the Majumdar-Ghosh ground state has perfect string order.    
However, it appears that, depending on which of the two degenerate ground 
states is actually realized, string order can be due to the spins along 
the rungs or to the spins along plaquette diagonals.

\vspace{.2in}
\vspace{.2in}

\vspace{.2in}
\vspace{.2in}

\breakon

Motivated by the above examples, similar to 
Refs.~\ref{white} and \ref{nishiyama}, we introduce two string order 
parameters   
\begin{eqnarray}
   {\cal O}^\alpha_{\rm odd} &=& - \lim_{|i-j| \rightarrow \infty} 
   \left\langle (S^{\alpha}_{i,1} + S^{\alpha}_{i,2})  
  \exp\left( i\pi \sum_{l=i+1}^{j-1} 
           (S^{\alpha}_{l,1} + S^{\alpha}_{l,2}) \right) 
   (S^{\alpha}_{j,1} + S^{\alpha}_{j,2})  \right\rangle  \, ,\nonumber \\
   {\cal O}^\alpha_{\rm even} &=& - \lim_{|i-j| \rightarrow \infty} 
   \left\langle (S^{\alpha}_{i+1,1} + S^{\alpha}_{i,2})  
  \exp\left( i\pi \sum_{l=i+1}^{j-1} 
           (S^{\alpha}_{l+1,1} + S^{\alpha}_{l,2}) \right) 
   (S^{\alpha}_{j+1,1} + S^{\alpha}_{j,2})  \right\rangle \, .
\label{eq:string_odd_even}   
\end{eqnarray}

\noindent
(The names ${\cal O}_{\rm odd}$ and ${\cal O}_{\rm even}$ will be made
clear below.) 
We saw that when one of the order parameters is finite, the other vanishes.  
So, the antiferromagnetic ladder has ${\cal O}_{\rm even} \neq 0$; the
composite spin model has ${\cal O}_{\rm odd} \neq 0$; the zig-zag ladder 
at the Majumdar-Ghosh point is special -- it can have either 
${\cal O}_{\rm even} \neq 0$ or ${\cal O}_{\rm odd} \neq 0$, depending 
on which of the two degenerate ground states is actually realized.  
However, it cannot have both finite simultaneously.

Let us now consider the topology of the VB's in the above examples; an
interesting pattern emerges. If we count the number 
of VB's crossing an arbitrary vertical line, we find that this number is 
always even for the ground state configurations of the antiferromagnetic 
ladder (Fig.~\ref{fig:configurationrvb}), while it is always odd for the 
diagonal ladder at $J_2 = J$ (Fig.~\ref{fig:configurationaklt}).
For the zig-zag ladder at the Majumdar-Ghosh point, the number of VB's 
crossing an arbitrary vertical line depends on which of the two degenerate 
ground states we consider: it is even for the state in 
Fig.~\ref{fig:majumdar}(a), while it is odd for the state in 
Fig.~\ref{fig:majumdar}(b).
\breakoff

\vspace{.2in}
\begin{figure}
\epsfxsize=3.25in
\centerline{\epsfbox{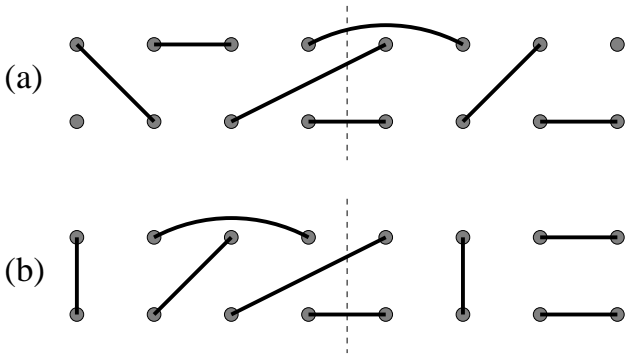} }
\vspace{.2in}
\caption{ Solid lines connecting sites represent VB's.  
 (a) $Q_y = odd$: the number of VB's crossing a vertical 
 line is odd. (b) $Q_y = even$: the number of VB's crossing a 
 vertical line is even.}
\label{fig:topology}
\end{figure}
\vspace{.2in}

These examples are special cases of a more general classification of 
SR-VB states. It has been shown\cite{bonesteel} that for SR-VB 
configurations on a two-dimensional square lattice, two topological 
numbers, $Q_x = even/odd$ and $Q_y = even/odd$, can be defined. 
They are determined by the parity of the number of SR-VB's crossing 
arbitrary horizontal and vertical lines parallel to the $x$- and $y$-axes, 
respectively. In the case of 2-leg ladders, only $Q_y$ is relevant; $Q_y$
can be either even or odd, as illustrated more generally in 
Fig.~\ref{fig:topology}.  Hence, there is a topological number which 
distinguishes between whether the number of SR-VB's cut by a vertical line 
is even ($Q_y = even$) or odd ($Q_y = odd$). For any finite size system, the 
even and odd sectors are coupled as long as there are VB's with length
comparable to the system size. However, when the system is gapped and 
thus has a SR-VB ground state, the tunneling amplitude between the two
sectors goes to zero exponentially fast as $L\to\infty$, and the ground 
state is a pure $Q_y = even$ or $Q_y = odd$ state in the thermodynamic
limit.  Note that in long-ranged VB ground states of gapless models, 
the even and odd sectors remain coupled; hence, no such topological 
distinction is possible.

It is also worth noting that for open boundary conditions 
 $Q_y = odd$ ground states have spin-1/2's localized at the ends of 
the ladder, while $Q_y = even$ states do not. As is obvious from 
Fig.~\ref{fig:topology}, these end spins occur for topological 
reasons. Their presence or absence is probably the simplest way 
to determine $Q_y$.  

From the above examples, the (topological) parity of the SR-VB 
ground state and the type of string order seem to be intimately 
related.  It appears that ground states with $Q_y = odd$ have 
${\cal O}_{\rm odd}$ string order, while ground states with 
 $Q_y = even$ have ${\cal O}_{\rm even}$ string order. 

Now suppose we smoothly vary the parameters of the Hamiltonian, such 
that we interpolate between models belonging to different topological 
classes; a phase transition necessarily occurs. A priori this transition 
could be either first order or second order, depending on the actual 
path in parameter space.  When the transition is second order, the 
string order parameters vanish at the transition point and the ground 
state becomes a long-ranged VB state.  In the next section we analyze 
this problem in the weak coupling limit using bosonization.


\section{Weak Coupling Analysis: Bosonization}

\begin{figure}
\epsfxsize=3.25in
\centerline{\epsfbox{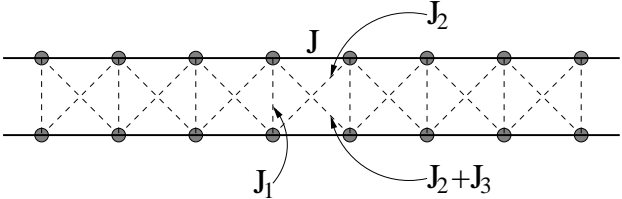} }
\vspace{.2in}
\caption{Most general ladder model which we will consider, 
 ${\cal H} = {\cal H}_0 + {\cal H}_1 + {\cal H}_2 + {\cal H}_3$.  }
\label{fig:laddergeneral}
\end{figure}

In the bosonization treatment of the ladder model shown in 
Fig.~\ref{fig:laddergeneral}, we start with two decoupled spin-1/2 chains 
and treat the interchain couplings perturbatively.  Our conventions, as well
as the bosonization of spin ladders, are presented in detail in 
Ref.~\ref{gene}.

\vspace{.2in}
\vspace{.2in}

The isotropic spin-1/2 Heisenberg chain is known to be critical.  The 
effective Hamiltonian for long wavelength excitations is 
\begin{equation}
 {\cal H}_{s=1/2} =  \frac{u}{2} \int dx 
     \left[ K \Pi^2  + \frac{1}{K} (\partial_x \Phi)^2 \right]
\end{equation}
where the bosonic phase field, $\Phi$, and its conjugate momentum,
$\Pi$, satisfy the commutation relation
\begin{equation} 
   [\Phi(x), \Pi(y)] = i\delta(x-y) \, .
\end{equation} 
For an isotropic antiferromagnetic spin chain, $K=1/2$. We will also
need the bosonized form of the spin operators.  They are\cite{gene}
\begin{eqnarray}
   S^{+}(x) & = & { S^{+}_j \over \sqrt{a}}  = \frac{ \exp\left( -i \sqrt{\pi} 
     \Theta  \right) } {\sqrt{2\pi a} } \left[ e^{-i(\pi x/a)} + 
      \cos(\sqrt{4\pi} \Phi ) \right] ,   \nonumber \\
   S^z(x) & = & { S^z_j  \over a } = \frac{1}{\sqrt{\pi}} \partial_x \Phi 
       + e^{i (\pi x/a)} \frac{ \sin( \sqrt{4\pi} \Phi) }{\pi a} \, ,
\label{spin-boson}
\end{eqnarray}
where the dual field, $\Theta$, is related to $\Pi$ by  
$\Pi = \partial_x \Theta$.

For the ladder, we simply attach a chain index to 
our fields.  Therefore,
\begin{eqnarray}
 {\cal H}_0 & = &  \frac{u}{2} \int dx 
   \left[ K \Pi_1^2  + \frac{1}{K} (\partial_x \Phi_1)^2 \right] \nonumber \\
    & &  + \frac{u}{2} \int dx  \left[ K \Pi_2^2  + \frac{1}{K} 
      (\partial_x \Phi_2)^2 \right]  \, .   
\end{eqnarray}
To bosonize the interchain coupling, we write the spin operators in terms 
of uniform and staggered components as
\begin{equation}
 {\bf S}_i(x) = {\bf J}_i(x) + (-1)^{x/a} {\bf n}_i(x) \, .
\end{equation}
We find
\begin{eqnarray}
 {\cal H}_1 & \sim & J_1 \int dx \left[ 
 {\bf J}_1(x) \cdot {\bf J}_2(x) + {\bf n}_1(x) \cdot {\bf n}_2(x)
  \right], \nonumber \\
 {\cal H}_2 & \sim & 2J_2 \int dx \left[ 
 {\bf J}_1(x) \cdot {\bf J}_2(x) - {\bf n}_1(x) \cdot {\bf n}_2(x)
  \right], \nonumber \\
 {\cal H}_3 & \sim & J_3 \int dx \left[ 
 {\bf J}_1(x) \cdot {\bf J}_2(x) - {\bf n}_1(x) \cdot {\bf n}_2(x)
  \right]. 
\label{interchain}
\end{eqnarray}
Inserting the expressions in Eq.~(\ref{spin-boson}) into 
Eq.~(\ref{interchain}) gives
\begin{eqnarray}
    {\cal H}_{i} & = &  \int {dx \over (2 \pi a)^2}  
   \left[ g_1^i \cos \left( \sqrt{4\pi} (\Phi_1 + \Phi_2) 
      \right)  \right. \nonumber \\ 
     & & \phantom{++} + g_2^i  \cos \left( \sqrt{4\pi} (\Phi_1 - \Phi_2) 
        \right) \nonumber \\  
     & & \phantom{++} + \left. g_3^i  \cos \left( \sqrt{\pi} (\Theta_1 - 
         \Theta_2) \right) \right]  \nonumber \\
     & &  + \frac{J_{\perp}^i}{\pi} \int dx 
          \partial_x \Phi_1 \partial_x \Phi_2   \nonumber \\
   & & + \int \frac{dx}{(2\pi a)^2} 
    \left[ g_4^i \cos \left( \sqrt{\pi} (\Theta_1 - \Theta_2) \right) \right.
     \nonumber \\
   & & \phantom{++} \times \cos \left( \sqrt{4\pi} (\Phi_1 + \Phi_2) \right)
     \nonumber \\
   & & \phantom{+} + g_5^i \cos \left( \sqrt{\pi} (\Theta_1 - \Theta_2) \right) 
       \nonumber \\
        & & \left. \phantom{++}  \times  \cos \left( \sqrt{4\pi} 
        (\Phi_1 - \Phi_2) \right)  \right] \, .
\end{eqnarray}
The $g_1^i$, $g_2^i$, and $g_3^i$ terms come from 
${\bf n}_1(x) \cdot {\bf n}_2(x)$, 
while the $J_{\perp}^i$, $g_4^i$, and $g_5^i$ terms come from 
${\bf J}_1(x) \cdot {\bf J}_2(x)$.   

For ${\cal H}_1$:
\begin{eqnarray}
 & & g_1^1 = -2J_1; \ \ \ g_2^1 = 2J_1; \ \ \ g_3^1 = 2\pi J_1, \nonumber \\
 & & J_{\perp}^1 = J_1; \ \ \  g_4^1 = \pi J_1; \ \ \ g_5^1 = \pi J_1 \, .
\end{eqnarray} 
For ${\cal H}_2$:
\begin{eqnarray}
 & & g_1^2 = 4J_2; \ \ \ g_2^2 = -4J_1; \ \ \ g_3^2 = -4\pi J_1, \nonumber \\
 & & J_{\perp}^2 = 2J_2; \ \ \  g_4^2 = 2\pi J_2; \ \ \ g_5^2 = 2\pi J_2 \, .
\end{eqnarray} 
For ${\cal H}_3$:
\begin{eqnarray}
 & & g_1^3 = 2J_3; \ \ \ g_2^3 = -2J_3; \ \ \ g_3^3 = -2\pi J_3, \nonumber \\
 & & J_{\perp}^3 = J_3; \ \ \  g_4^3 = \pi J_3; \ \ \ g_5^3 = \pi J_3 \, .
\end{eqnarray} 

It is useful to define the fields 
\begin{equation}
 \Phi_{s,a} = \frac{1}{\sqrt{2}} \left( \Phi_1 \pm \Phi_2 \right) \, ,  \ \ \ 
 \Theta_{s,a} = \frac{1}{\sqrt{2}} \left( \Theta_1 \pm \Theta_2 \right) \,.
\end{equation}
In terms of these fields our Hamiltonian,
${\cal H} = {\cal H}_0 + {\cal H}_1 + {\cal H}_2 + {\cal H}_3$,
is 
\begin{eqnarray}
  {\cal H}  & = &  \frac{u_s}{2} \int dx \left[ K_s \Pi_s^2  + 
         \frac{1}{K_s} (\partial_x \Phi_s)^2 \right] \nonumber \\
      & & +  g_1 \int {dx \over (2 \pi a)^2} 
         \cos \left( \sqrt{8\pi} \Phi_s \right)   \nonumber \\
    & & +  \frac{u_a}{2} \int dx  \left[ K_a \Pi_a^2  +  
         \frac{1}{K_a} (\partial_x \Phi_a)^2 \right]  \nonumber \\
    & & + \int {dx \over (2 \pi a)^2} 
    \left[ g_2 \cos  \left( \sqrt{8\pi} \Phi_a \right) 
     +  g_3 \cos \left( \sqrt{2\pi} \Theta_a \right) \right] \nonumber \\
    & & + g_4 \int \frac{dx}{(2\pi a)^2} 
    \cos \left( \sqrt{2\pi} \Theta_a \right)
    \cos \left( \sqrt{8\pi} \Phi_s \right)   \nonumber \\
    & & + g_5 \int \frac{dx}{(2\pi a)^2} 
    \cos \left( \sqrt{2\pi} \Theta_a \right)
    \cos \left( \sqrt{8\pi} \Phi_a \right)   \,,
\end{eqnarray}
where 
\begin{eqnarray}
g_1 & = & -2J_1 + 4J_2 + 2J_3 \ \ \ ,  \ \ \ g_2 = 2J_1 -4J_2 - 2J_3 \, , 
  \nonumber \\
g_3 & = & 2\pi J_1 - 4\pi J_2 - 2\pi J_3 \ \ \ , \ \ \  
g_4 = \pi J_1 + 2\pi J_2 + \pi J_3 \, , \nonumber \\
& & \hspace{.4in} g_5 = \pi J_1 + 2\pi J_2 + \pi J_3 \, , \nonumber \\
& & \hspace{.4in} J_{\perp} = J_1 + 2J_2 + J_3 \, .
\end{eqnarray}
Also,
\begin{eqnarray}
 K_s = K \left( 1 + \frac{KJ_{\perp}}{u \pi} \right)^{-1/2} \,,  
  &  & \ \ \ 
 u_s = u \left( 1 + \frac{K J_{\perp}}{u \pi} \right)^{1/2} \,, \nonumber \\
    & & \\
 K_a = K \left( 1 - \frac{KJ_{\perp}}{u \pi} \right)^{-1/2} \, ,
    & & \ \ \ 
 u_a = u \left( 1 - \frac{KJ_{\perp}}{u \pi} \right)^{1/2} \, . \nonumber 
\end{eqnarray}
For $J_{\perp} \ll 1$ we have
\begin{eqnarray}
 K_s \approx K \left( 1 - \frac{K J_{\perp}}{2u \pi} \right),
 & &  \ \
 u_s \approx u \left( 1 + \frac{K J_{\perp}}{2u \pi} \right) , 
 \nonumber \\
 K_a \approx K \left( 1 + \frac{K J_{\perp}}{2u \pi} \right),
  &  &  \ \
 u_a \approx u \left( 1 - \frac{K J_{\perp}}{2u \pi} \right) .
\end{eqnarray}

We are interested in whether or not the interchain coupling causes 
a gap in the excitation spectrum.  Therefore,
we would like to identify the relevant operators;
these operators will ``pin'' their arguments,
thus causing gaps to appear.  To do this we consider the scaling
dimensions of the operators in the interchain coupling.\cite{tsvelik,gene}  
The scaling dimensions of the operators are the following: 
$\left[\cos \left( \sqrt{8\pi} \Phi_s \right)\right] = 2K_s$; 
$\left[\cos \left( \sqrt{8\pi} \Phi_a \right)\right] = 2K_a$; 
$\left[\cos \left( \sqrt{2\pi} \Theta_a \right)\right] = 1/(2K_a)$; 
$\left[\cos \left( \sqrt{2\pi} \Theta_a \right) 
 \cos \left( \sqrt{8\pi} \Phi_s \right)\right] = 2K_s + 1/(2K_a)$;
$\left[\cos \left( \sqrt{2\pi} \Theta_a \right) 
 \cos \left( \sqrt{8\pi} \Phi_a \right)\right] = 2K_a + 1/(2K_a)$.
Therefore, $g_1$ will grow at large distances for $K_s < 1$; 
$g_2$ will grow for $K_a < 1$; $g_3$ will grow for $K_a > 1/4$;  
$g_4$ will grow for $2K_s + 1/(2K_a) < 2$; $g_5$ will grow for 
$2K_a + 1/(2K_a) < 2$.

In what follows, we will consider the phases and transitions that
occur when we vary $J_1$, $J_2$, and $J_3$.  In order to make things
more tractable, we will consider two-dimensional slices in the full 
$J_1-J_2-J_3$ space.

\vspace{.2in}
\begin{figure}
\epsfxsize=3.0in
\centerline{\epsfbox{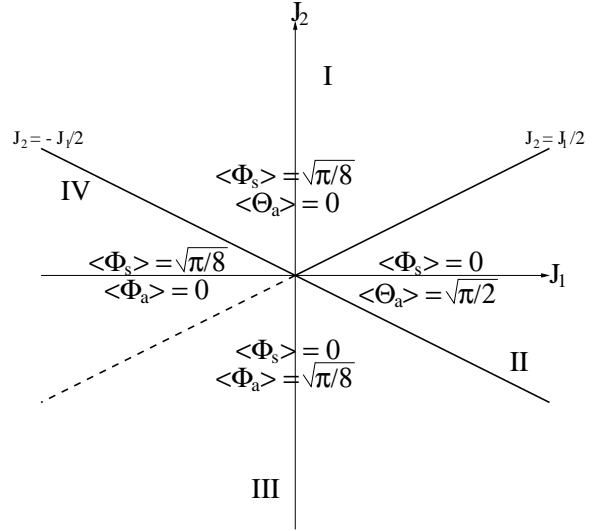} }
\caption{Phase diagram for $J_1 \neq 0$ and $J_2 \neq 0$ with $J_3 = 0$.}
\label{fig:J1J2phases}
\end{figure}
\vspace{-0.0in}


\subsection{$J_1 \neq 0$ and $J_2 \neq 0$ with $J_3 = 0$} 

In this case, for $J_2 = 0$ and $J_1 = J$ we recover
the usual antiferromagnetic ladder; for $J_2 = 0$ and 
$J_1 \rightarrow -\infty$ we recover a spin-1 chain, where 
the spins on each rung form an $S=1$. Similarly, for $J_1 = 0$ and
$J_2 = J$ we recover the composite spin representation for 
a spin-1 chain, and it was previously shown that the composite spin 
representation has the same low energy physics as the true spin-1
chain.\cite{composite}

The phase diagram in the $J_1 - J_2$ plane is shown in 
Fig.~\ref{fig:J1J2phases}.  
In region $I$ the $g_1$ and $g_3$ terms are the most relevant. 
Therefore, $\Phi_s$ and $\Theta_a$ are pinned with  
$\langle\Phi_s\rangle = (2n+1)\sqrt{\pi/8}$ 
and $\langle\Theta_a\rangle = m\sqrt{2\pi}$.  Without loss of generality,
we can choose $n=0$ and $m=0$.  This gives 
$\langle\Phi_s\rangle = \sqrt{\pi/8}$ and $\langle\Theta_a\rangle = 0$.  
In region $II$ the $g_1$ and $g_3$ terms are again the most relevant.
Again, $\Phi_s$ and $\Theta_a$ are pinned.  Now, however,  
$\langle\Phi_s\rangle = n\sqrt{\pi/2}$ and 
$\langle\Theta_a\rangle = (2m+1)\sqrt{\pi/2}$.  Choosing $n=0$ and $m=0$,
$\langle\Phi_s\rangle = 0$ and $\langle\Theta_a\rangle = \sqrt{\pi/2}$.  
In region $III$ the $g_1$ and $g_2$ terms are the most relevant.
Therefore, $\Phi_s$ and $\Phi_a$ are pinned with 
$\langle\Phi_s\rangle = n\sqrt{\pi/2}$ and 
$\langle\Phi_a\rangle = (2m+1)\sqrt{\pi/8}$.  Choosing $n=0$ and $m=0$
gives $\langle\Phi_s\rangle = 0$ and $\langle\Phi_a\rangle = \sqrt{\pi/8}$.  
In region $IV$, similar to region $III$, the $g_1$ and $g_2$ terms are
the most relevant, so $\Phi_s$ and $\Phi_a$ are pinned.  
However, in this region $\langle\Phi_s\rangle = (2n+1)\sqrt{\pi/8}$ and 
$\langle\Phi_a\rangle = m\sqrt{\pi/2}$.  Choosing $n=0$ and $m=0$,
$\langle\Phi_s\rangle = \sqrt{\pi/8}$ and $\langle\Phi_a\rangle = 0$.  
There are also two special lines in the phase diagram.  Along the line
$J_2 = J_1/2$, the ${\bf n}_1 \cdot {\bf n}_2$ terms vanish and only the 
${\bf J}_1 \cdot {\bf J}_2$ terms remain.  For $J_1,J_2 < 0$, the system 
is gapless; for $J_1,J_2 > 0$, the $g_4$ term is marginally relevant and 
the system is gapped.  However, the ground state is two-fold degenerate
(for $J_1,J_2 > 0$):
$\langle \Phi_s \rangle = (2n+1)\sqrt{\pi/8}$, 
$\langle \Theta_a \rangle = 2m \sqrt{\pi/2}$ 
or $\langle \Phi_s \rangle = 2n\sqrt{\pi/8}$, 
$\langle \Theta_a \rangle = (2m+1) \sqrt{\pi/2}$.  
Choosing $n=0$ and $m=0$, we have  
$\langle \Phi_s \rangle = \sqrt{\pi/8}$, $\langle \Theta_a \rangle = 0$ 
or $\langle \Phi_s \rangle = 0$, $\langle \Theta_a \rangle = \sqrt{\pi/2}$.
The other special line is $J_2=-J_1/2$. Along this line the 
${\bf J}_1 \cdot {\bf J}_2$ terms vanish and the 
${\bf n}_1 \cdot {\bf n}_2$ terms all have the same scaling dimension = 1.
Along this line the spectrum is gapped.  This line will be discussed in 
greater detail in Sec.~V.


\vspace{.6in}
\begin{figure}
\epsfxsize=2.75in
\centerline{\epsfbox{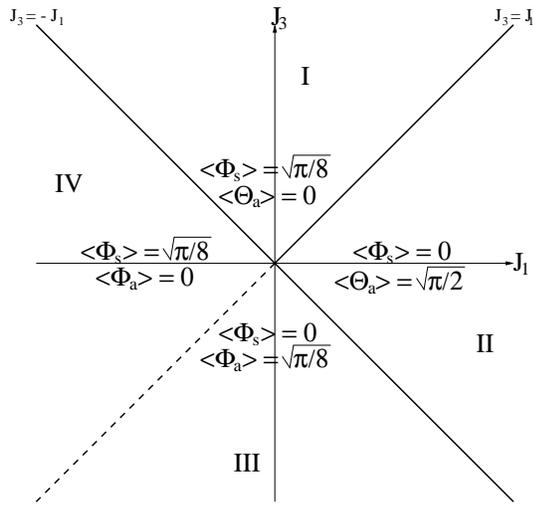} }
\caption{Phase diagram for $J_1 \neq 0$ and $J_3 \neq 0$ with $J_2 = 0$.}
\label{fig:J1J3phases}
\end{figure}

\subsection{$J_1 \neq 0$ and $J_3 \neq 0$ with $J_2 = 0$} 

In this case, for $J_3 = 0$ and $J_1 = J$ we 
recover the usual antiferromagnetic ladder; for $J_3 = 0$ 
and $J_1 \rightarrow -\infty$ we recover a true spin-1 chain.
For $J_1 = J_3 \neq 0$ we have a zig-zag ladder; in particular, 
for $J_1 = J_3 = 2J$ we have the Majumdar-Ghosh point where 
the ground state is dimerized with two-fold degeneracy.

The phase diagram in the $J_1 - J_3$ plane is shown in 
Fig.~\ref{fig:J1J3phases}.  The regions $I$, $II$, $III$, and $IV$
have properties identical to the $J_1 - J_2$ phase diagram discussed 
above.  The line $J_3 = -J_1$ has properties identical to the line
$J_2 = -J_1/2$ discussed above, and will be discussed in greater 
detail in Sec.~V.  
The line $J_1 = J_3$ is 
special.  As pointed out by Nersesyan {\em et al.},\cite{nersesyan} we 
must be careful of the ``twist'' operators which appear.  Along this line, 
the interchain coupling of the staggered components can be written as
\begin{equation}
 H_{{\rm int}} \sim \int dx \ \left[ {\bf n}_1 \partial_x {\bf n}_2 
                   - {\bf n}_2 \partial_x {\bf n}_1 \right] .
\end{equation}
Explicitly, the terms are
\begin{eqnarray}
 H_{{\rm int}} & = & g'_1 \int {dx \over (2\pi a)^2} 
    \partial_x \Phi_a \sin \left( \sqrt{8\pi} \Phi_s \right)   \nonumber \\
 & + & g'_2 \int {dx \over (2\pi a)^2} 
    \partial_x \Phi_s \sin  \left( \sqrt{8\pi} \Phi_a \right)
    \nonumber \\ 
 & + & g'_3 \int {dx \over (2\pi a)^2} 
    \partial_x \Theta_s \sin \left( \sqrt{2\pi} \Theta_a \right)  \, ,
\end{eqnarray}
where $g'_1$, $g'_2$, $g'_3 \sim J_3$.
These terms are subtle because they have non-zero conformal spin.  As pointed
out in Ref.\ \ref{tsvelik}, a seemingly irrelevant operator with non-zero 
conformal spin can generate relevant operators.  However, since we are only 
considering the SU(2) symmetric case, the terms generated are less relevant 
than the ${\bf J}_1 \cdot {\bf J}_2$ terms already present. Therefore, similar 
to the line $J_2 = J_1/2$, for $J_3 < 0$ the system is gapless; for $J_3 > 0$ 
the $g_4$ term is marginally relevant and the spectrum is gapped, with the 
ground state being two-fold degenerate.


\section{Discussion of the Results}

As discussed in Sec.~III, the two-leg ladder models we have considered 
can have either
${\cal O}^{\alpha}_{\rm odd}$ or ${\cal O}^{\alpha}_{\rm even}$ string
order, but not both simultaneously.  In our case, since we are only 
considering SU(2) symmetric models, 
${\cal O}^x = {\cal O}^y = {\cal O}^z$.  Therefore, for simplicity, 
we focus only on ${\cal O}^z$.     

To bosonize the string order parameter, we first write it in
a more convenient form. Using the identity 
$\exp\left( i\pi S^z \right) = 2i S^z$, we can write

\breakon
\begin{eqnarray}
 {\cal O}^{z}_{\rm odd} & = &  \frac{1}{4}\lim_{|i-j| \rightarrow \infty} 
   \left\langle \left( e^{i\pi S^{\alpha}_{i,1}} 
                     + e^{i\pi S^{\alpha}_{i,2}} \right) 
  \exp\left( i\pi \sum_{l=i+1}^{j-1} 
           (S^{\alpha}_{l,1} + S^{\alpha}_{l,2}) \right) 
 \left( e^{i\pi S^{\alpha}_{j,1}}  
      + e^{i\pi S^{\alpha}_{j,2}} \right) 
  \right\rangle  \, ,  \nonumber \\
 {\cal O}^{z}_{\rm even} & = & \frac{1}{4} \lim_{|i-j| \rightarrow \infty} 
   \left\langle \left( e^{i\pi S^{\alpha}_{i+1,1}}  
                     + e^{i\pi S^{\alpha}_{i,2}} \right) 
  \exp\left( i\pi \sum_{l=i+1}^{j-1} 
           (S^{\alpha}_{l+1,1} + S^{\alpha}_{l,2}) \right) 
 \left( e^{i\pi S^{\alpha}_{j+1,1}}  
      + e^{i\pi S^{\alpha}_{j,2}} \right) 
  \right\rangle  \, .  
\end{eqnarray}
\breakoff

\noindent   
Bosonizing ${\cal O}^{z}_{\rm odd}$ and ${\cal O}^{z}_{\rm even}$
gives 
\begin{equation}
 {\cal O}^{z}_{{\rm even/odd}} = \lim_{|x-y| \rightarrow \infty} 
 \left\langle e^{i\sqrt{2\pi} \Phi_s(x)} e^{-i\sqrt{2\pi} \Phi_s(y)}    
 \right\rangle \, .
\end{equation} 
We see that all we need is for $\Phi_s$ to get pinned to have string order.
The operators for both ${\cal O}^{z}_{\rm odd}$ and ${\cal O}^{z}_{\rm even}$ 
have the same bosonized form because the nonlocal string operator makes 
the continuum limit insensitive to physics occurring on the order of a
single lattice spacing, such as whether triplets lie predominantly along 
rungs or along diagonals.  Therefore, the bosonized string order parameter 
tells us that we have string order, but it does not tell us in which 
topological sector the order exists.  Taking into account the physical
picture we get from the VB states in Sec.~III, we can understand the 
various regions and transition lines which have been obtained in the phase 
diagrams by bosonization.  The results are summarized in 
Table~\ref{table:results}.   

\end{multicols}

\begin{table}
\caption{Results for the $J_1-J_2$ and $J_1-J_3$ phase diagrams.} 
\begin{center}
\begin{tabular}{ccccc}
  & \hspace{.5in}  $I$   
  & \hspace{.5in} $II$  
  & \hspace{.5in} $III$ 
  & \hspace{.5in} $IV$  \\
\tableline \\
 $\Phi_s$  &  \hspace{.5in}   $\langle \Phi_s \rangle = \sqrt{\pi/8}$  
           &  \hspace{.5in}  $\langle \Phi_s \rangle = 0$ 
           &  \hspace{.5in}  $\langle \Phi_s \rangle = 0$  
           &  \hspace{.5in}  $\langle \Phi_s \rangle = \sqrt{\pi/8}$  \\  \\
 $\Phi_a$, $\Theta_a$  
 & \hspace{.5in} $\langle \Theta_a \rangle = 0$ 
 & \hspace{.5in}  $\langle \Theta_a \rangle = \sqrt{\pi/2}$  
 & \hspace{.5in} $\langle \Phi_a \rangle = \sqrt{\pi/8}$
 & \hspace{.5in} $\langle \Phi_a \rangle = 0$     \\  \\
 Order Parameter &  \hspace{.5in}  ${\cal O}_{\rm odd}$  
                 &  \hspace{.5in} ${\cal O}_{\rm even}$ 
                 &  \hspace{.5in} ${\cal O}_{\rm even}$  
                 &  \hspace{.5in} ${\cal O}_{\rm odd}$    \\  \\
\tableline   \\
\end{tabular}
\begin{tabular}{ccc}
  \hspace{.1in} $J_2 = J_1/2$:  &  \hspace{.1in} $J_2,J_3 > 0$ --- 
                                 first order transition 
                 &  \hspace{.1in} $J_2,J_3 < 0$ --- second order 
                                             transition \\
  \hspace{.1in} $(J_3 = J_1$)   & \hspace{.1in} 
         $\langle \Phi_s \rangle = \sqrt{\pi/8}$ , 
         $\langle \Theta_a \rangle = 0 $  or
         $\langle \Phi_s \rangle = 0$ , 
         $\langle \Theta_a \rangle = \sqrt{\pi/2}$ 
  & \hspace{.1in}  $\Phi_s$ and $\Phi_a$ critical \\ \\
 \hspace{.1in} $J_2 = -J_1/2$: &  \hspace{.1in} 
                                 level crossing in the excited states
                 &  \hspace{.1in}       \\
 \hspace{.1in} $(J_3 = - J_1)$ & \hspace{.1in} & \hspace{.1in}  \\  \\
\tableline
\end{tabular}
\end{center}
\label{table:results}
\end{table}

\begin{multicols}{2}

\subsection{$J_1 \neq 0$ and $J_2 \neq 0$ with $J_3 = 0$}    

The line $J_1 = 0$ with $J_2 > 0$ is continuously related to the 
composite spin model; the composite spin model has ${\cal O}_{\rm odd}$ 
string order.  Therefore, it appears that region $I$ is continuously 
related to this model, and hence has ${\cal O}_{\rm odd}$ string order.  
The line $J_2 = 0$ with $J_1 > 0$ is continuously related to the usual
antiferromagnetic ladder; the antiferromagnetic ladder has 
${\cal O}_{\rm even}$ string order.  Therefore, it appears that region $II$
is continuously related to this model, and hence has ${\cal O}_{\rm even}$ 
string order.  Along the line $J_1 = 0$ with $J_2 < 0$, we have
ferromagnetic interchain coupling along plaquette diagonals.  For
$|J_2| \approx J_1$
the ground state is similar to the RVB state
of the ladder with antiferromagnetic interchain coupling.  Hence, this
model has ${\cal O}_{\rm even}$ string order.  It appears that
region $III$ is continuously related to this model, and hence has
${\cal O}_{\rm even}$ string order.  Finally, the line $J_2 = 0$ with 
$J_1 < 0$ is continuously related to the spin-1 chain in which the spins 
on a rung form an effective $S=1$.  Since the ground state of the spin-1
chain is described by the AKLT state, this model has ${\cal O}_{\rm odd}$
string order.  Therefore, it appears that region $IV$ is continuously 
related to this model, and hence has ${\cal O}_{\rm odd}$ string order.  
We see that a transition between ${\cal O}_{\rm odd}$ and 
${\cal O}_{\rm even}$ string order occurs along the line $J_2 = J_1/2$.  
For $J_2 < 0$ the transition is second order; for $J_2 > 0$, there is a 
marginally relevant operator which drives the transition first order.
The line $J_2 = -J_1 / 2$ is interesting, so we discuss it in detail.

Along the line $J_2 = -J_1/2$ there is a change in the properties 
of the system: above the line, $\Theta_a$ is pinned; below the line,
$\Phi_a$ is pinned.  However, we believe this is a level crossing in
the excited states; the properties of the ground state remain the
same.  Hence, the system does not undergo a phase transition
when we cross this line.  To show this, it is useful to express 
${\cal H}_a$ in terms of Majorana fermions.  We begin on the line
$J_2 = -J_1/2$; along this line, the ${\bf J}_1 \cdot {\bf J}_2$ terms
vanish and the ${\bf n}_1 \cdot {\bf n}_2$ terms all have the same scaling 
dimension.  Rescaling our fields,
\begin{equation}
 \Pi_a \rightarrow \frac{1}{\sqrt{K_a}} \Pi_a \ \ \ , \ \ \ 
 \Phi_a \rightarrow \sqrt{K_a} \Phi_a  \, ,
\end{equation}
${\cal H}_a$ has the form
\begin{eqnarray}
  & & \hspace{.45in} {\cal H}_a  = \frac{u_a}{2} \int dx  \left[ \Pi_a^2  +  
         (\partial_x \Phi_a)^2 \right]  \nonumber \\
    & &  + \int {dx \over (2 \pi a)^2} 
    \left[ g_2 \cos  \left( \sqrt{4\pi} \Phi_a \right) 
     +  g_3 \cos \left( \sqrt{4\pi} \Theta_a \right) \right]  \, .
\end{eqnarray}
Using that\cite{tsvelik} 
\begin{eqnarray}
  & & \frac{u}{2} \int dx \left[ \Pi^2  + 
         (\partial_x \Phi)^2 \right]   \nonumber \\      
  & & \hspace{.4in} =  -i u \int dx \left[ 
         \psi_R^{\dagger} \partial_x \psi_R - 
         \psi_L^{\dagger} \partial_x \psi_L \right] \, ,  
 \nonumber \\
  & & \frac{1}{\pi a} \cos (\sqrt{4\pi} \Phi) 
  = i\left( \psi_R^{\dagger} \psi_L - 
            \psi_L^{\dagger} \psi_R \right)  \, ,  \\
  & & \frac{1}{\pi a} \cos (\sqrt{4\pi} \Theta) 
  = i\left( \psi_R^{\dagger} \psi_L^{\dagger} - 
            \psi_L^{\dagger} \psi_R^{\dagger} \right)  \, ,
 \nonumber
\end{eqnarray}
${\cal H}_a$ can be written as
\begin{eqnarray}
 {\cal H}_a & = &  -i u_a \int dx \left( 
         \psi_{a,R}^{\dagger} \partial_x \psi_{a,R} - 
         \psi_{a,L}^{\dagger} \partial_x \psi_{a,L} \right)
 \nonumber \\
  & + & \frac{i}{2} \int \frac{dx}{2\pi a} \left[
        g_2 \left( \psi_{a,R}^{\dagger} \psi_{a,L} - 
        \psi_{a,L}^{\dagger} \psi_{a,R} \right) \right.
 \nonumber \\  & &  \hspace{.2in} \left.
    + g_3 \left( \psi_{a,R}^{\dagger} \psi_{a,L}^{\dagger} - 
        \psi_{a,L}^{\dagger} \psi_{a,R}^{\dagger} \right) \right].
\end{eqnarray}
Now introduce two independent Majorana fermions, $\xi$ and $\eta$, 
defined by
\begin{equation}
 \psi_{a,R}  = \frac{1}{\sqrt{2}} (\xi_R + i \eta_R)
 \ \ \ , \ \ \ 
 \psi_{a,L}  = \frac{1}{\sqrt{2}} (\xi_L + i \eta_L) \, .
\end{equation}
Finally, ${\cal H}_a$ can be written as
\begin{eqnarray}
 {\cal H}_a & = & \frac{-i v_a}{2} \int dx
       \left[ ( \xi_R \partial_x \xi_R -
                \xi_L \partial_x \xi_L)  \right.
  \nonumber \\   
      & & \hspace{.85in} + \left. ( \eta_R \partial_x \eta_R -
              \eta_L \partial_x \eta_L) \right] 
  \nonumber\\
      & + & \frac{i}{2} \int \frac{dx}{2\pi a} 
           \left[ (g_2 - g_3) \xi_R \xi_L +
                  (g_2 + g_3) \eta_R \eta_L \right].
\end{eqnarray}
This is the Hamiltonian for two massive Majorana fermions.  As is 
well known, massive Majorana fermions describe the long distance 
properties of the Ising model away from criticality.

For $J_2 \approx -J_1/2$ (i.e. $J_2 = -J_1/2 + \delta$, 
$|\delta| \ll 1$), the ${\bf J}_1 \cdot {\bf J}_2$ terms do not 
vanish.  However, very close to the line $J_2 = -J_1 / 2$, we can still 
write ${\cal H}_a$ in terms of Majorana fermions.  The 
${\bf J}_1 \cdot {\bf J}_2$ terms can be written as four-fermion
interactions which just renormalize the velocity and fermion masses 
\cite{shelton}.  The key thing to notice is that when we cross the line 
$J_2 = -J_1 /2 $, the values of the fermion masses change, but their 
{\it signs} do {\it not} change.  It is well known that the (Majorana)
fermion mass changing sign corresponds to the order-disorder transition 
of the Ising model.  Since there is no change in sign when we cross the 
line $J_2 = -J_1 / 2$, the structure of the ground state does not appear
to change.  Therefore, we interpret the change from $\Theta_a$ being 
pinned to $\Phi_a$ being pinned as a level crossing in the excited states.  
Hence, the system does not appear to undergo a phase transition when 
crossing this line.


\subsection{$J_1 \neq 0$ and $J_3 \neq 0$ 
            with $J_2 = 0$} 
The line $J_1 > 0$ with $J_3 = 0$ is continuously related to the 
usual antiferromagnetic ladder; the antiferromagnetic ladder has 
${\cal O}_{\rm even}$ string order.  Therefore, it appears that region 
$II$ is continuously related to this model, and hence has 
${\cal O}_{\rm even}$ string order.  The line $J_1 < 0$ with $J_3 = 0$ 
is continuously related to the spin-1 chain in which the spins
on a rung form an effective $S=1$.  Since the ground state of the 
spin-1 chain is described by the AKLT state, this model has
${\cal O}_{\rm odd}$ string order.  Therefore it appears that 
region $IV$ is continuously related to this model, and hence has 
${\cal O}_{\rm odd}$ string order.  Along the line $J_1 =0$, the 
coupling along the rungs is zero and only the diagonal interchain 
coupling, $J_3$, is nonzero.  This is similar to the case when only 
$J_1 \neq 0$, except with chain-1 shifted to the right by one lattice 
spacing.  Therefore, $J_3 > 0$ is similar to the usual antiferromagnetic
ladder and $J_3 < 0$ is similar to the spin-1 chain, except with chain-1 
shifted to the right by one lattice spacing.  Hence, $J_3 > 0$ has 
${\cal O}_{\rm odd}$ string order, and $J_3 < 0$ has 
${\cal O}_{\rm even}$ string order.  It appears that region $I$ 
is continuously related to the line $J_3 > 0$ and that region $III$ 
is continuously related to the line $J_3 < 0$.  Therefore, region $I$ 
has ${\cal O}_{\rm odd}$ string order and region $III$ has 
${\cal O}_{\rm even}$ string order.  A phase transition occurs along 
the line $J_3 = J_1$.  For $J_3 < 0$ the transition is second 
order, while for $J_3 > 0$ a marginally relevant operator drives
the transition first order.  Similar to the 
line $J_2 = -J_1/2$, the system changes
character when crossing the line $J_3 = -J_1$.  Above the line,
$\Theta_a$ is pinned; below the line $\Phi_a$, is pinned.  
Similarly, we believe that there is no phase transition as we cross 
this line; it is a level crossing in the excited states.

It is interesting to note that in our model, the zig-zag ladder 
(i.e. the line $J_3 = J_1$) is actually a transition line.  For 
$J_3 > 0$ the line is a first order transition line, while for 
$J_3 < 0$ the transition is second order.  Therefore, the 
Majumdar-Ghosh point actually lies on a first order transition line.


\section{Concluding Remarks}

\begin{figure}
\epsfxsize=3.0in
\centerline{\epsfbox{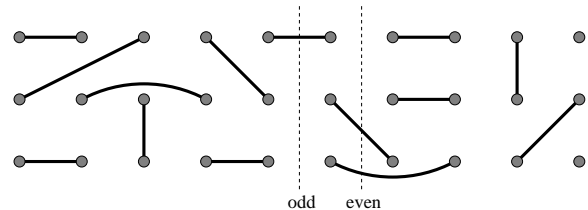} }
\vspace{.2in}
\caption{A VB configuration which could occur for a three-leg
ladder, provided the couplings are chosen so that the ground state is
described by a SR-VB state.  Notice that the number of bonds
crossing a vertical line alternates.}
\label{fig:evenodd}
\end{figure}

In this paper we studied the gapped phases in two-leg spin ladders.  
The ground states of these ladders are well described by SR-VB states.  
There are two topologically distinct classes characterized by whether 
the number of VB's cut by a vertical line is even ($Q_y = even$) or 
odd ($Q_y = odd$).  Note that this classification of $Q_y = even$ 
and $Q_y = odd$ can be used for even-leg ladders but not for odd-leg 
ladders.  For odd-leg ladders, one gets an even-odd alternation, as 
shown schematically in Fig.~\ref{fig:evenodd}.  This even-odd alternation 
implies a two-fold degenerate ground state, consistent with the 
Lieb-Schultz-Mattis theorem.\cite{bonesteel}

\vspace{.2in}
\vspace{.2in}

Associated with $Q_y = even$ and $Q_y = odd$, we considered the ``even'' 
and ``odd'' string order parameters of Eq.~\ref{eq:string_odd_even} for 
the ladder model shown in Fig.~\ref{fig:laddergeneral}. Using known results 
for particular values of the coupling constants along with bosonization, 
we obtained the phase diagrams in the $J_1 - J_2$ and $J_1 - J_3$ planes, 
shown in Figs.~\ref{fig:J1J2phases} and \ref{fig:J1J3phases}, respectively. 
While these results cover only parts of the $J_1-J_2-J_3$ parameter space,
we believe that the association of the string order parameters
${\cal O}^{\alpha}_{\rm even}$ and ${\cal O}^{\alpha}_{\rm odd}$ with
$Q_y = even$ and $Q_y = odd$ is appropriate for this model in general.

We should emphasize that this classification of $Q_y = even$ and
 $Q_y = odd$ relies on the possibility of writing the singlet ground 
state of the ladder as a superposition of VB configurations.  This 
may break down in anisotropic models, and so by introducing anisotropic 
couplings the two topological sectors may be coupled.

Another interesting problem is related to the spin-1 chain with bilinear 
and biquadratic exchange interactions\cite{fath} and spin ladders with
4-spin plaquette couplings,\cite{solyom,nersesyan2} both having
a non-Haldane-like dimerized phase.  Although $Q_y$ is still a good 
topological number for the dimerized ground state, it is not clear if 
string order is simply due to the short-ranged nature of the VB's 
and survives the transition from the Haldane phase to
the dimerized phase.

It is clear from this analysis that the apparently featureless spin liquid
phase of spin-gapped two-leg ladders actually has a rich underlying 
topological structure. It remains to be seen what role these ideas may
play in the doped systems.  More precisely, does this topological structure 
survive when the system is doped, and is pairing ultimately related to the 
topological structure?


\section*{ Acknowledgements }

We would like to thank L. Balents and G. Sierra for helpful
discussions.  EHK gratefully acknowledges the warm hospitality of 
Argonne National Laboratory, where parts of this manuscript were 
written. GF and JS are grateful to the University of California at 
Santa Barbara for their warm hospitality.  This work was supported by 
the Joint US-Hungarian Grant No. 555, the Hungarian Research Fund
(OTKA) grant No. 30173 (GF and JS), and DOE grant No. 85-ER45197 
(EHK and DJS).



\end{multicols}
\end{document}